\documentclass[12pt]{article}
\usepackage{epsf,fancyhdr}
\usepackage{amssymb}
\usepackage[dvips]{graphicx}
\usepackage{graphics}
\usepackage{epsfig}
\usepackage{latexsym}
\usepackage[figuresright]{rotating}
\def\slash#1{\not\!#1}
\begin{document}
\begin{center}
{\Large \textbf{Recursive technique for evaluation of Feynman
diagrams }}
\\
\textbf{\large V. V. Andreev} \footnote{\textsf{E-mail}:\textbf{
ANDREEV@GSU.UNIBEL.BY}}\\
Gomel State University, Physics Department, \\
Gomel, Belarus
\end{center}
\begin{abstract}
A method is presented in which matrix elements for some processes
are calculated recursively. This recursive calculational
technique is based on the method of basis spinors.
\end{abstract}

\section{Introduction}
The possibility of investigation higher and higher energies at
present and future colliders, entails the necessity of predicting
and calculating with high precision more and more complicated
processes. When the number of final particles is high it becomes
hard, even to calculate the corresponding tree level Feynman
diagrams  and the final expression of cross section is often an
intricated function of several variables, inadequate for
practical use. This has lead to that it is necessary abandon the
standard methods for perturbative calculations and to use instead
the new effective ones.

The standard method to obtain a cross section with the fermions in
perturbative quantum field theories is to reduce the squared
amplitude to a trace from products of $\gamma$-matrices.  An
alternative approach is to calculate the Feynman amplitudes
directly.  The idea of calculating amplitudes has a long enough
history. In 1949  it was suggested in Ref.\cite{Powell} to
calculate a matrix element by means of explicit form of
$\gamma$-matrices and Dirac spinors (more detailed bibliography
on the problem can be found in Refs.~\cite{Galynski,Bondarev}).

Various methods of calculating the reaction amplitudes with
fermions have been proposed and successfully applied in recent
years. In general the methods of matrix element calculation can be
classified into two basic types. The first type includes methods
of direct numerical calculation of the Feynman diagrams. The
second type includes methods of analytical calculations of
amplitudes with the subsequent numerical calculations of cross
sections. Notice that there are methods of calculating cross
sections without the Feynman diagrams \cite{Caravaglios,Helac}.

Analytical methods of calculating the Feynman amplitudes  can be
divided into two basic groups. The first group involves the
analytical methods that reduce the calculation of  $S$-matrix
element to a trace calculation. The reduction of a matrix element
to trace calculation from products of $\gamma$-matrices underlies
a large number of methods (see, Refs.\cite{Galynski,Bondarev} and
Refs. \cite{Bellomo}-\cite{Vega} etc.). In this method the matrix
element is expressed as the algebraic function in terms of scalar
products of four-vectors and their contractions with the
Levi-Civita tensor.

The second group involves the analytical methods that practically
do not use the operations with traces from products of
$\gamma$-matrices. The method of the CALCUL group which was used
for the calculations of the reactions with massless fermions is
the most famous among \cite{Berendz}-\cite{Giele}. The basic idea
behind the CALCUL method is to replace $S$-matrix element by
spinor products of bispinors and to use the fact that expressions
$\bar{u}_{\lambda}\left(p\right)u_{-\lambda}\left(k\right)$ are
simple scalar functions of the momenta $p,k$  and the helicity
$\lambda$. However, the operation of matrix element reduction is
not so simple as the calculation of traces. It requires the use
of Chisholm spinor identities (see \cite{Kleiss}). Also it takes
the representation of contraction $ \slash{p} = p^\mu \gamma_\mu$
with four-momenta $p^\mu$ and polarization vectors of external
photons through bispinors. For gauge massive bosons the
additional mathematical constructions are needed \cite{Kleiss}.

There are generalizations of the CALCUL method for massive Dirac
particles both for special choices of the fermion polarization
(\cite{Kleiss},\cite{Berendz3}-\cite{Hagiwara}) and for an
arbitrary fermion polarization \cite{Gongora,Andreev}. We call
the polarization states of fermions in Ref.\cite{Berendz3,Kleiss}
as Berends-Daverveldt-Kleiss-Stirling or $BDKS$-states.

Notice that for massless fermions we can obtain amplitude in
terms of the scalar products of four-momentum vectors and
current-like constructions of the type $J^{\mu} \sim
\bar{u}_{\lambda}\left(p\right) \gamma^{\mu} u_ {\lambda} \left
(k\right)$. The components of $J^{\mu}$ are calculated by means of
momentum components $p,k$ (so called $E$-vector formalism; see
Ref. \cite{Papadopoulos}).

For $BDKS$-states Ref.\cite{Ballestrero} presents the iterative
scheme of calculation that reduces expression for the fermion
chain $\bar{u}_{\lambda} \left(p\right)Q\; u_{\lambda}
\left(k\right) $ to the combination of spinor products
$\bar{u}_{\lambda}\left(p\right)u_{\lambda}\left(k\right)$ and
(or) $\bar{u}_{\lambda}\left(p\right)\gamma^{\mu}\left(g_V+g_A
\gamma_5 \right) u_{\lambda} \left(k\right)$ by means of inserting
the complete set of  non-physical states of bispinors (with
$p^2<0$) into the fermion chain.

In all  above-mentioned methods the spinor products and current
constructions were calculated by means of traces and then used as
scalar functions of the momenta and of the helicities (similar to
scalar product of four-vectors).

Due to their easy implementation  methods of matrix element
calculation have become a basis for modern programs dealing with
evaluation cross-sections for various processes. Examples for such
programs are generators \texttt{HELAS} \cite{Murayama},
\texttt{GRACE} \cite{Grace}, \texttt{MadGraph} \cite{MadGraph},
\texttt{O'MEGA} \cite{omega}, \texttt{FeynArts/FormCalc4}
\cite{FeynArts,Hahn} and \texttt{CompHEP} \cite{Comphep} (planned
the calculation of the matrix element \cite{Ilyin}). There is
large number of more specialized programs as \texttt{AMEGIC++}
\cite{Amegic}, \texttt{ALPGEN} \cite{Alpgen}, \texttt{WPHACT}
\cite{wphact}, \texttt{ LUSIFER }\cite{lusifer} and \textit{et
al}. The detailed list of such programs can be found in
Ref.\cite{Harlander}.

In the  paper we describe  an approach  to  Feynman diagrams which
is based on the using of an isotropic tetrad in Minkowski space
and basis spinors connected with it (see
\cite{Andreev3b},\cite{Andreev1}). Here we don't use  an explicit
form of Dirac spinors and $\gamma$--matrices and the operation of
trace calculations. The method is based on the active using of the
massless basis spinors connected with isotropic tetrad vectors
and we will call it as Method of Basis Spinors (\texttt{MBS}).

In this method as well as in the trace methods the matrix element
of Feynman amplitudes is reduced to the combination of scalar
products of momenta and polarization vectors. Unlike spinor
technique in different variants \cite{Berendz}-\cite{Kleiss1},
this method doesn't use either Chisholm identities, or the
presentation of the contraction $\slash{p}$ with four vector $p$
and of the polarization vector of bosons through the bispinors.
Unlike \texttt{WvD} spinor technique \cite{Giele},\cite{Dittmaier}
\texttt{MBS} doesn't use special Feynman rules for calculating of
the matrix elements.

We propose to use recursion relations as a technique to evaluate
the Feynman amplitudes of processes. The advantage of recursive
technique is that for calculation of a $n+1$ matrix element of
some process we can use the calculation of $n$ process. Both for
analytic and numerical evaluation this is asset.

\section{Method of Basis Spinors}

When evaluating a Feynman amplitude involving the fermions, the
amplitude is expressed as the sum of terms that have the form
\begin{eqnarray}
&& \mathcal{M}_{\lambda _p,\lambda _k}\left(p,s_p\;,\; k,\;s_k\;
;Q\right)=
\nonumber\\
&& =\mathcal{M}_{\lambda _p,\lambda _k}\left(\left[p\right],
\left[k\right] ;Q\right)=\bar{u}_{\lambda _p}\left( p,s_p\right)
Q ~u_{\lambda _k}\left(k,s_k\right)\;, \label{anpic1}
\end{eqnarray}
where $\lambda_{p}$ and $\lambda_{k}$ are the polarizations  of
the external particles with four-momentum $p,k$ and arbitrary
polarization vectors $s_p,s_k$. The operator $Q$ is the sum of
products of Dirac $\gamma$-matrices.  The matrix element
(\ref{anpic1}) with Dirac spinors is a scalar function. Thus, it
should be expressible in terms of scalar functions formed from
the spin and momentum four-vectors of the fermions, including $p,
s_p, k, s_k$ and  the operator $Q$.

We will now consider that in the our approach this matrix element
(\ref{anpic1}) can be represented as linear combinations of the
products of the lower-order matrix elements.

\subsection{Isotropic tetrad}

We use the metric and matrix convention as in the book by Bjorken
and Drell \cite{Bjorken1}, i.e. the Levi-Civita tensor is
determined as $\epsilon_{0 1 2 3}=1$ and the matrix $ \gamma_5=i
\gamma ^{0} \gamma ^{1} \gamma ^{2} \gamma ^{3}$. Let us
introduce the orthonormal four-vector basis in Minkowski space
which satisfies the relations:
\begin{equation}
\label{pic1} l_0^{\mu} \cdot l_0^{\nu} -l_1^{\mu} \cdot l_1^{\nu}
-l_2^{\mu} \cdot l_2^{\nu}-l_3^{\mu} \cdot l_3^{\nu} = g^{\mu
\nu}, ~~\left(l_{A} \cdot l_{B}\right)=g_{A B},
\end{equation}
where $g$ is the Lorentz metric tensor.

With the help of the basis vectors $l_{A}\left(A=0,1,2,3\right)$
we can define lightlike vectors, which form the isotropic tetrad
in Minkowski space (see, \cite{Borodulin})
\begin{equation}
b_\rho =\frac{l_0+\rho l_3}{2}\;,\; n_\lambda
=\frac{\lambda\;l_1+\mathrm{i} l_2}{2}\;, ( \rho ,\lambda =\pm
1)\;. \label{pic2}
\end{equation}
From Eqs. (\ref{pic1}), (\ref{pic2}) it follows  that
\begin{equation}
(b_\rho \cdot b_{-\lambda })=\frac{\delta _{\lambda,\;\rho
}}{2}~,~~(n_\lambda \cdot n_{-\rho })=\frac{\delta
_{\lambda,\;\rho }}{2}~,~~\left(b_\rho \cdot n_\lambda \right)
=0\;, \label{pic3}
\end{equation}
\begin{equation}
g^{\mu \nu}=2\sum_{\lambda =-1}^1\left[ b_\lambda ^\mu \cdot
b_{-\lambda}^\nu +n_\lambda ^\mu \cdot n_{-\lambda }^\nu
\right]\;. \label{pic4}
\end{equation}
It is always possible to construct the basis of an isotropic
tetrad (\ref{pic2}) as numerical four-vectors
\begin{equation}\label{pic4a}
\left(b_{\pm 1}\right)_{\mu}=\left(1/2\right)\left\{1, 0, 0, \pm
1\right\}\;,\; \left(n_{\pm
1}\right)_{\mu}=\left(1/2\right)\left\{0, \pm 1, \mathrm{i},
0\right\}
\end{equation}
or by means of the physical vectors for reaction.

For practical applications it is convenient to introduce
additional four-vectors
\begin{equation}
\tilde{b}_\rho =2\; b_\rho ,\; \tilde{n}_\lambda=2\; n_{\lambda}
\label{add1}
\end{equation}

By means of the isotropic tetrad vectors we can determine the
polarization vectors of massless (also  and massive) vector
bosons. For photons with momentum $k^\mu$ and helicity
$\lambda=\pm 1$ we use the following definition of polarizations
in the axial gauge
\begin{eqnarray}
&& \varepsilon _{\lambda }\left( k\right) =\frac{\left(k \cdot
\tilde{n}_{-\lambda}\right) \tilde{b}_{-1}}{\sqrt{2}\;\left(k
\cdot \tilde{b}_{-1}\right)}-\frac{\tilde{n}_{-\lambda}}{\sqrt{2}}
\label{picc1b}
\end{eqnarray}
provided that, the four-vectors $k, b_1, b_{-1}$ are linearly
independent.

\subsection{Massless basis spinors} \label{sec:level2}

By means of the isotropic tetrad vectors (\ref{pic2}) we define
{\it massless basis spinors} $u_{\lambda}\left(b_{-1}\right)$ and
$u_{\lambda}\left(b_{1}\right)$
\begin{equation}
\slash{b}_{-1} u_\lambda \left( b_{-1}\right) =0\;, ~~u_\lambda
\left(b_{1}\right) \equiv
\slash{b}_{1}u_{-\lambda}\left(b_{-1}\right)\;,\label{pic7}
\end{equation}
\begin{equation}
\omega _\lambda u_\lambda \left(b_{\pm 1}\right)= u_\lambda
\left(b_{\pm 1}\right) \label{pic8}
\end{equation}
with the matrix $\omega _{\lambda} = 1/2 \hskip 1pt \left(
1+\lambda \gamma_5\right)$ and the normalization condition
\begin{equation}\label{pic8a}
u_\lambda \left( b_{\pm 1}\right) \bar{u}_\lambda \left(b_{\pm
1}\right) =\omega _\lambda \slash{b}_{\pm 1}.
\end{equation}

The relative phase between basis spinors with different helicity
is given by
\begin{equation}
\slash{n}_\lambda u_{-\nu }\left( b_{-1}\right) =\delta_{\lambda,
\nu} u_\lambda \left( b_{-1}\right). \label{pic9}
\end{equation}

The important property of basis spinors (\ref{pic7}) is the
completeness relation
\begin{equation}
\sum_{\lambda,A=-1}^{1} u_\lambda \left( b_A\right)
\bar{u}_{-\lambda} \left(b_{-A}\right)= I\;, \label{pic10}
\end{equation}
which follows from Eqs.(\ref{pic7}), (\ref{pic9}). Thus, the
arbitrary Dirac spinor can be decomposed in terms of basis spinors
$u_{\lambda}\left(b_{A} \right)$.

\subsection{Dirac spinors and basis spinors}

Arbitrary  Dirac spinor can be determined through the basis spinor
(\ref{pic7}) with the help of projection operators
$u_{\lambda_{p}} \left(p,s_p\right)\bar{u}_{\lambda_{p}}
\left(p,s_p\right)$.

The Dirac spinors $w^A_\lambda \left(p,s_p\right)$ for massive
fermion and antifermion with four-momentum $p\;( p^2=m_p^2 )$ ,
arbitrary polarization vector $s_p$ and spin number $\lambda= \pm
1$  can be obtained with the help of basis spinors by means of
equation:
\begin{eqnarray}
&& w^A_\lambda \left(p,s_p\right) =\frac{\left(\slash{p}+A
m_p\right) \left(1+ \lambda
\gamma_{5}\slash{s}_{p}\right)}{2\sqrt{\left(b_{-1}\cdot \left(
p+m_p s_p \right)\right)}} u_{-A \times \lambda} \left(b_{-1}
\right)\nonumber\\
&& =\frac{ \left[ \slash{\xi}^{p}_{1}+A\;
\slash{\xi}^{p}_{-1}\slash{\xi}^{p}_{1}/ m_p \right]
{u}_{-A\times \lambda_{p}} \left(b_{-1}\right)}{\sqrt{\left(
\tilde{b}_{-1}\cdot \xi^{p}_{1}\right)}}\nonumber\\
&& =T_{\lambda}\left(p,s_p\right) u_{-A \times \lambda}
\left(b_{-1} \right)\;. \label{anpic8}
\end{eqnarray}
The notation $w^{A}_{\lambda _p}\left(p,s_p\right)$ stands for
either $u_{\lambda _p}\left( p,s_p\right)$ (bispinor of fermion;
$A=+1$) or $\upsilon_{\lambda _p}\left( p,s_p\right)$ (bispinor of
antifermion; $A=-1$). Here we have introduced the abbreviations
\begin{equation}
\xi _{\pm 1}^p =\frac{p \pm m_p s_p}{2}\;. \label{anpic11}
\end{equation}

The bispinors $u_\lambda \left( p,s_p\right)$ and
$\upsilon_\lambda \left( p,s_p\right) $ satisfy Dirac equations
and spin conditions for massive fermion and antifermion
\begin{eqnarray}
\slash{p} \hskip 2pt u_\lambda \left(p,s_p\right)& =&m_p \hskip
2pt u_{\lambda }\left(p,s_p\right), \hskip 14pt \slash{p} \hskip
2pt \upsilon_\lambda \left(p,s_p\right) = - m_p \hskip 2pt
\upsilon_{\lambda }\left(p,s_p\right) , \nonumber\\ \gamma_5
\slash{s}_{p} \; u_\lambda \left(p,s_p\right) &=& \hskip 2pt
\lambda \; u_{\lambda }\left(p,s_p\right),~ \gamma_5
\slash{s}_{p} \hskip 2pt \upsilon_\lambda \left(p,s_p\right) =
\lambda \hskip 2pt \upsilon_{\lambda }\left(p,s_p\right).
\label{anpic8a}
\end{eqnarray}

We also found, that the Dirac spinors of fermions and
antifermions are related by
\begin{equation}
\upsilon_\lambda \left( p,s_p\right)= -\lambda \gamma
_5\;u_{-\lambda }\left( p,s_p\right), \hskip 7pt
\bar{\upsilon}_\lambda \left( p,s_p\right) =\bar{u}_{-\lambda
}\left( p,s_p\right) \lambda \hskip 2pt \gamma_5\;. \label{stp31}
\end{equation}

Let us consider the particular case of Eq.(\ref{anpic8})--the
$BDKS$ polarization states of fermions, as they are the most-used
in calculations of matrix elements. The polarization vector of
$BDKS$-states is defined as follows
\cite{Kleiss,Berendz3,Andreev,Ballestrero}:
\begin{equation}
s_{KS} \equiv s_p= \frac{p}{m_p}-\frac{m_p\;b_{-1}}{\left(p \cdot
b_{-1}\right)}\; . \label{stp33aa}
\end{equation}

Performing a calculation for Eqs.(\ref{pic7}),(\ref{anpic8}) and
Eq.(\ref{stp33aa}), we find the simple result for massive Dirac
spinor \cite{Kleiss, Ballestrero,Andreev}:
\begin{equation}
\label{stp34aa}
w_{\lambda}^{A} \left(p,s_{KS}\right) = \frac{\left(\slash{p}+A\;
m_p \right)\;u_{-A\times\lambda
}\left(b_{-1}\right)}{\sqrt{\left(p \cdot \tilde{b}_{-1}\right)}}
\end{equation}
Notice that, in Ref.\cite{Ballestrero} relation between the Dirac
spinor of fermion and the Dirac spinor of antifermion differs
from the Eq.(\ref{stp31}).

The Dirac spinor $u_\lambda \left(p\right)$ of massless fermion
with momentum $p$ ($p^2=0, \left(p \cdot b_{-1}\right) \not =
0$)  and helicity $\lambda$ is defined by (for example, see
Ref.\cite{Kleiss})
\begin{equation}
u_\lambda \left(p\right) = \frac{\slash{p}\; u_{-\lambda
}\left(b_{-1}\right)}{\sqrt{\left(p \cdot \tilde{b}_{-1}\right)
}}\; . \label{anpic6}
\end{equation}

\subsection{Main equations of \texttt{MBS}}

The spinor products of massless basis spinors (\ref{pic7}) are
determined by
\begin{equation}
\bar{u}_\lambda \left( b_C\right) u_{\rho}\left( b_A\right) =
\delta_{\lambda, -\rho}\; \delta_{C, -A},\;~\;C,A=\pm 1,\;
~\lambda,\rho=\pm 1\;. \label{stp18}
\end{equation}

With the help of Eq.(\ref{pic4}) Dirac matrix $\gamma^\mu $ can
be rewritten as
\begin{equation}
\gamma ^\mu =\sum_{\lambda =-1}^1\left[\slash{b}_{-\lambda}
\tilde{b}_\lambda ^\mu +\slash{n}_{-\lambda } \tilde{n}_\lambda
^\mu \right]\;. \label{pic5}
\end{equation}

Using the Eqs.(\ref{pic7}),(\ref{pic8}) and (\ref{pic9}) we can
obtain that
\begin{equation}
\gamma^\mu u_\lambda \left( b_A\right) =\tilde{b}_A^\mu
\;u_{-\lambda }\left(b_{-A}\right) -A \;{\tilde{n}}_{-A \times
\lambda }^\mu \; u_{-\lambda }\left(b_A\right) \label{pic11}
\end{equation}
and
\begin{equation}
\label{pic13a} \gamma_{5}\; u_\rho \left(b_A\right)= \rho \;u_\rho
\left(b_A\right)\;.
\end{equation}
Eqs.(\ref{pic11})-(\ref{pic13a}) and Eq.(\ref{stp18}) underlie
the method of basis spinors (\texttt{MBS})
\cite{Andreev3b,Andreev1}.

By means of the Eq.(\ref{pic11}) we can determine that product of
two $\gamma$-matrices can be represented as
\begin{equation}
\gamma ^\mu \gamma ^\nu u_\lambda \left( b_A\right)=Y_{A,\;\lambda
}^{\mu,\; \nu}\;u_{\lambda }\left(b_{A}\right) -A \;
{X}_{A,\;\lambda }^{\mu, \nu}\; u_{\lambda }\left(b_{-A}\right)\;,
\label{pic14a}
\end{equation}
where  $X^{\mu, \nu}, Y^{\mu, \nu}$ are the Lorentz tensors:
\begin{eqnarray}
&& X_{A,\;\lambda}^{\mu,\; \nu}=\tilde{b}_{A}^{\mu}\cdot
\tilde{n}_{-A\times \lambda}^{\nu}- \tilde{n}_{-A\times
\lambda}^{\mu}\cdot \tilde{b}_{A}^{\nu}\;, \label{anpic12new1}\\
&& Y_{A,\;\lambda}^{\mu,\; \nu}=\tilde{b}_{-A}^{\mu} \cdot
\tilde{b}_{A}^{\nu}+ \tilde{n}_{A\times \lambda}^{\mu} \cdot
\tilde{n}_{-A\times\lambda}^{\nu}\;. \label{anpic12new2}
\end{eqnarray}

The  product  $\mathcal{S}^{n}=\gamma ^{\mu_{1}} \gamma
^{\mu_{2}}\ldots \gamma ^{\mu_{n}}$ can be written as
\begin{eqnarray}
&&\mathcal{S}^{n}\;u_\lambda \left( b_A\right)
\nonumber\\
&& =\mathcal{B}_{A,\;\lambda}^{\left\{\mu_{1},\ldots
\mu_{n}\right\}}\;u_{\lambda_{n}^{\prime}
}\left(b_{A_{n}^{\prime}}\right) -A\;
\mathcal{N}_{A,\;\lambda}^{\left\{\mu_{1},\ldots
\mu_{n}\right\}}\; u_{\lambda_{n}^{\prime}
}\left(b_{-A_{n}^{\prime}}\right)\;, \label{pic14ab}
\end{eqnarray}
where
\begin{equation}
\lambda_{n}^{\prime}=\left(-1\right)^{n}\lambda\;,\;
A_{n}^{\prime}=\left(-1\right)^{n} A \label{al}
\end{equation}
and $\mathcal{B}_{A,\;\lambda}^{\left\{\mu_{1},\ldots
\mu_{n}\right\}}$,
$\mathcal{N}_{A,\;\lambda}^{\left\{\mu_{1},\ldots
\mu_{n}\right\}}$ are some Lorentz tensors, which are related with
isotropic tetrad vectors (\ref{add1}).

As follows from Eqs.(\ref{pic11}),(\ref{pic14a}) we have that in
particular cases:
\begin{eqnarray}
&&
\mathcal{B}_{A,\;\lambda}^{\left\{\mu_{1}\right\}}=\tilde{b}_{A}^{\mu_{1}},\;
\mathcal{N}_{A,\;\lambda}^{\left\{\mu_{1}\right\}}=\tilde{n}_{-A\times\lambda}^{\mu_{1}}\;,
\label{bn1}\\
&& \mathcal{B}_{A,\;\lambda}^{\left\{\mu_{1},\;
\mu_{2}\right\}}=Y_{A,\;\lambda}^{\mu_{1},\; \mu_{2}} \;,\;
\mathcal{N}_{A,\;\lambda}^{\left\{\mu_{1},\;\mu_{2}\right\}}=X_{A,\;\lambda}^{\mu_{1},\;
\mu_{2}} \;. \label{bn2}
\end{eqnarray}

\section{Recursion relations for the matrix elements}

The \textbf{basic idea of Method of Basis Spinors} is to replace
Dirac spinors in Eq.(\ref{anpic1}) by massless basis spinors
(\ref{pic7}), and to use the Eq.(\ref{stp18})and
Eqs.(\ref{pic11})-(\ref{pic13a}) for calculation of matrix
element (\ref{anpic1}) in terms of scalar functions
$\mathcal{B},\mathcal{N}$. With the help of the Eq.(\ref{anpic8})
the matrix element (\ref{anpic1}) transforms to fermion ``string''
with massless basis spinors $u_{\lambda}\left(b_{A}\right)$ i.e.
\begin{eqnarray}
&& \mathcal{M}_{\lambda _p,\lambda _k}\left(p,s_p\;,\; k,s_k\;
;Q\right)=
\nonumber\\
&& =\bar{u}_{-\lambda _p} \left(b_{-1}\right)
T_{\lambda_{p}}\left(p,s_p\right)Q\;
T_{\lambda_{k}}\left(k,s_k\right)u_{-\lambda
_k}\left(b_{-1}\right)=\nonumber\\
&& =\mathcal{M}_{-\lambda _p,-\lambda _k}\left(b_{-1},\;b_{-1}\;
;T_{\lambda_{p}}\left(p,s_p\right)Q\;
T_{\lambda_{k}}\left(k,s_k\right)\right)\;, \label{anpic9a}
\end{eqnarray}
where operator $T_{\lambda}$ is determined by Eq.(\ref{anpic8}).
Let us consider a special variants of a matrix element.

\subsection{Basic matrix element}
Let us consider an important type of a matrix element
(\ref{anpic1}), when $p=b_{-C}$ and $k=b_A$, i.e.
\begin{eqnarray}
&& \mathcal{M}_{-\sigma,\rho}\left(b_{-C}\;,\;b_A\; ;
Q\right)\equiv \Gamma^{C,\;A}_{\sigma ,\rho}\left[Q\right]
\nonumber\\
&&  = \bar{u}_{-\sigma} \left( b_{-C}\right) Q \;u_\rho \left(
b_A\right)\;. \label{pic13}
\end{eqnarray}
We call this type of matrix element as \textbf{basic matrix
element}.

Note that, the matrix element (\ref{anpic1}) is a particular case
of basic matrix element i.e.
\begin{equation}
\mathcal{M}_{\lambda _p,\lambda _k}\left(p,s_p,\; k,s_k\;
;Q\right)=\Gamma^{1,-1}_{\lambda _p ,-\lambda
_k}\left[T_{\lambda_{p}}\left(p,s_p\right)Q\;
T_{\lambda_{k}}\left(k,s_k\right)\right] \;.\label{anpic9ab}
\end{equation}

With the help of the completeness relation (\ref{pic10}) we can
obtain the recursion formula for $\Gamma^{C,A}_{ \sigma
,\rho}\left[Q_1 Q_2\right]$
\begin{equation}
\label{pic14} \Gamma^{C,\;A}_{\sigma ,\rho}\left[Q_1 Q_2\right]=
\sum_{D,\lambda=-1}^{1}\Gamma^{C,\;D}_{\sigma, \lambda
}\left[Q_1\right] \Gamma^{D,\;A}_{\lambda, \rho}\left[Q_2\right].
\end{equation}

By means of the relations
(\ref{pic11}),(\ref{pic14a}),(\ref{pic14ab}) and Eq.(\ref{stp18})
it is easy to calculate $\Gamma^{C,A}_{\sigma, \rho}$ in terms of
the isotropic tetrad vectors. For instance,
\begin{eqnarray}
&& \Gamma^{C,\;A}_{\sigma,\;\rho
}\left[\gamma^{\mu}\right]=\delta_{\sigma,\; -\rho}
\left(\delta_{C,\;-A}\; \tilde{b}_{A}^{\mu}-A\;\delta_{C,A} \;
\tilde{n}_{-A \times \rho }^{\mu}\right)\;,
\label{pic16}\\
&& \Gamma^{C,\;A}_{\sigma,\;
\rho}\left[\gamma^\mu\;\gamma^\nu\right]= \delta_{\sigma , \rho}
\left(\delta_{C,A} Y_{A,\;\rho }^{\mu,\; \nu}-A\;\delta_{C,-A}
{X}_{A,\;\rho }^{\mu,\; \nu} \right)\; \label{pic16a}
\end{eqnarray}
and
\begin{eqnarray}
&& \Gamma^{C,\;A}_{\sigma,\;\rho}\left[\gamma ^{\mu_{1}} \gamma
^{\mu_{2}}\ldots \gamma ^{\mu_{n}}\right]=
\Gamma^{C,\;A}_{\sigma,\; \rho}\left[\mathcal{S}^{n}\right]
\nonumber\\
&& =\delta_{\sigma,\;\rho_{n}^{\prime} } \left(\delta_{C,\;
A_{n}^{\prime}}\; \mathcal{B}_{A, \;\rho}^{\left\{\mu_{1},\ldots
\mu_{n}\right\}} -A\;\delta_{C, -A_{n}^{\prime}}\;
\mathcal{N}_{A,\;\rho}^{\left\{\mu_{1},\ldots
\mu_{n}\right\}}\right)\;. \label{pic16ab}
\end{eqnarray}

With the help of the Eqs. (\ref{pic14}) and (\ref{pic16ab}) we
obtain recursion relations for $\mathcal{B}_{A,
\lambda}^{\left\{\mu_{1},\ldots \mu_{n}\right\}}$ and
$\mathcal{N}_{A, \lambda}^{\left\{\mu_{1},\ldots \mu_{n}\right\}}$
\begin{eqnarray}
&& \mathcal{B}_{A,\; \lambda}^{\left\{\mu_{1},\ldots
\mu_{n}\right\}}=\mathcal{B}_{A_{n-k}^{\prime},\;
\lambda_{n-k}^{\prime}}^{\left\{\mu_{1},\ldots
\mu_{k}\right\}}\mathcal{B}_{A,
\;\lambda}^{\left\{\mu_{k+1},\ldots
\mu_{n}\right\}}+\left(-1\right)^{n-k+1}\mathcal{N}_{-A_{n-k}^{\prime},\;
\lambda_{n-k}^{\prime}}^{\left\{\mu_{1},\ldots
\mu_{k}\right\}}\mathcal{N}_{A,\;
\lambda}^{\left\{\mu_{k+1},\ldots \mu_{n}\right\}}\;,
\label{recurbn1}\\
&&  \mathcal{N}_{A,\; \lambda}^{\left\{\mu_{1},\ldots
\mu_{n}\right\}}=\mathcal{B}_{-A_{n-k}^{\prime},\;
\lambda_{n-k}^{\prime}}^{\left\{\mu_{1},\ldots
\mu_{k}\right\}}\mathcal{N}_{A,\;
\lambda}^{\left\{\mu_{k+1},\ldots
\mu_{n}\right\}}+\left(-1\right)^{n-k}\mathcal{N}_{A_{n-k}^{\prime},\;
\lambda_{n-k}^{\prime}}^{\left\{\mu_{1},\ldots
\mu_{k}\right\}}\mathcal{B}_{A,\;
\lambda}^{\left\{\mu_{k+1},\ldots \mu_{n}\right\}}
\;.\label{recurbn2}
\end{eqnarray}

The recursion Eqs.(\ref{recurbn1})-(\ref{recurbn2}) allow to
convert scalar functions $\mathcal{B},\mathcal{N}$ into Lorentz
tensors in terms of isotropic tetrad vector with the help of
Eq.(\ref{bn1}) or Eq.(\ref{bn2}). For example, the constructions $
q_1^{\mu_{1}}  q_2^{\mu_{2}}  q_3^{\mu_{3}} \mathcal{B}_{A,\;
\lambda}^{\left\{\mu_{1},\;\mu_{2},\;\mu_{3}\right\}}=\mathcal{B}_{A,\;
\lambda}^{\left\{q_1,\;q_2,\;q_3\right\}}$ and $q_1^{\mu_{1}}
q_2^{\mu_{2}}  q_3^{\mu_{3}} \mathcal{N}_{A,
\lambda}^{\left\{\mu_{1},\;\mu_{2},\;\mu_{3}\right\}}=\mathcal{N}_{A,\;
\lambda}^{\left\{q_1, \;q_2,\; q_3\right\}}$ can be represented
as a combination of the scalar functions $X,\;Y$ and scalar
products of the isotropic tetrad vectors:
\begin{eqnarray}
&& q_1^{\mu_{1}}  q_2^{\mu_{2}}  q_3^{\mu_{3}} \mathcal{B}_{A,\;
\lambda}^{\left\{\mu_{1},\;\mu_{2},\;\mu_{3}\right\}}=\mathcal{B}_{A,\;
\lambda}^{\left\{q_1,\; q_2,\; q_3\right\}} \nonumber\\
&&=\left(q_3 \cdot \tilde{b}_{A}\right)Y_{-A,\;-\lambda}^{q_{1},\;
q_{2}}+ \left(q_3 \cdot \tilde{n}_{-A \times
\lambda}\right)X_{A,\;-\lambda}^{q_{1},\; q_{2}}\;,  \label{bn3}
\end{eqnarray}
\begin{eqnarray}
&& q_1^{\mu_{1}}  q_2^{\mu_{2}}  q_3^{\mu_{3}} \mathcal{N}_{A,\;
\lambda}^{\left\{\mu_{1},\;\mu_{2},\;\mu_{3}\right\}}=\mathcal{N}_{A,\;
\lambda}^{\left\{q_1,\; q_2,\; q_3\right\}} \nonumber\\
&&=\left(q_3 \cdot \tilde{n}_{-A
\times\lambda}\right)Y_{A,\;-\lambda}^{q_{1},\; q_{2}}- \left(q_3
\cdot \tilde{b}_{A}\right)X_{-A,\;-\lambda}^{q_{1},\; q_{2}}\;.
\label{bn4}
\end{eqnarray}

\subsection{Decomposition coefficients}

The next type of lower-order matrix element (\ref{anpic1}) is
\begin{eqnarray}
&& \mathcal{M}_{\rho,\;\lambda _p}\left(b_{A}\;,
\left[p\right];I\right)\equiv \mathcal{M}_{\rho,\;\lambda
_p}\left(b_{A}\;,\left[p\right]\right)=
\nonumber\\
&&= \bar{u}_{\rho} \left(b_{A}\right)u_{\lambda_{p}}
\left(p,s_p\right)\;. \label{anpic3}
\end{eqnarray}
The matrix element (\ref{anpic3}) is determined by the
decomposition coefficients of an arbitrary Dirac spinor
$u_{\lambda_{p}} \left(p,s_p\right)$ on basis spinors
(\ref{pic7}).

With the help of the Eqs.(\ref{anpic8}),(\ref{anpic6}) the matrix
element (\ref{anpic3})  transforms to
\begin{eqnarray}
&& \mathcal{M}_{\rho,\;\lambda _p}\left(b_A,\left[p\right]\right)=
\nonumber\\
&& =\frac{\bar{u}_{\rho} \left(b_{A}\right)\left[
\slash{\xi}^{p}_{1}+ \slash{\xi}^{p}_{-1}\slash{\xi}^{p}_{1}/ m_p
\right] \bar{u}_{-\lambda_{p}} \left(b_{-1}\right)}{\sqrt{\left(
\tilde{b}_{-1}\cdot \xi^{p}_{1}\right)}}
 \label{anpic1aa}
\end{eqnarray}
for the massive fermions with arbitrary vector of polarization and
transforms to
\begin{eqnarray}
&& \mathcal{M}_{\rho,\;\lambda
_p}\left(b_A,p\right)=\frac{\bar{u}_{\rho}\left(b_{A}\right)
\slash{p} \;u_{-\lambda
_p}\left(b_{-1}\right)}{\sqrt{\left(\tilde{b}_{-1}\cdot p\right)}}
\label{anpic1aa1}
\end{eqnarray}
for massless fermions.

Using the Eqs.(\ref{pic11})-(\ref{pic13a}) and (\ref{pic14a}) the
matrix element (\ref{anpic3}) is reduced to an algebraic
expression in terms of scalar products of isotropic tetrad
vectors and physical vectors  or in terms of components of
four-vectors.

Let us consider massless fermions. Using
Eqs.(\ref{stp18}),(\ref{pic11}) we obtain, that
\begin{equation}
\mathcal{M}_{\rho,\;\lambda}\left(b_A,p\right)=\delta_{\lambda,-\rho}
\left(\delta_{A,-1}\sqrt{\left(p \cdot \tilde{b}_{-1}\right)}+
\delta_{A,1}\frac{\left(p \cdot
\tilde{n}_{-\lambda}\right)}{\sqrt{\left(p \cdot
\tilde{b}_{-1}\right)}}\right)\;. \label{anpic7}
\end{equation}
For numerical calculations, as well as in the  case of spinor
techniques, it is convenient to determine the (\ref{anpic7})
through the momentum components $p =\left (p^0\right.$, $\;p^x =
p^0 \sin\theta_p\sin\varphi_p$, $p^y= p^0
 \sin\theta_p\cos\varphi_p$, $\left. p^z=p^0 \cos\theta_p\right)$
\begin{eqnarray}
&& \mathcal{M}_{\rho,\;\lambda}\left(b_A,\;p\right)=
\delta_{\lambda,-\rho}\left[ \delta_{A,-1}\sqrt{ p^{-}}-
\delta_{A,1}\;\lambda \exp\left(-\mathrm{i} \lambda
\varphi_{p}\right)\sqrt{p^{+}}\right]=
\nonumber\\
&&=\delta_{\lambda,-\rho}\sqrt{2 p_0}\left[ \delta_{A,-1}\sin
\frac{\theta_p}{2}- \delta_{A,1}\;\lambda\; \cos
\frac{\theta_p}{2}\exp\left(-\mathrm{i} \lambda \varphi_{p}\right)
\right], \label{stp24}
\end{eqnarray}
where
$$
p^{\pm}=p^0\pm p^z, ~~ p^x+\mathrm{i} \lambda p^y=\sqrt{\left(p^x
\right)^2+\left(p^y \right)^2} \exp\left(\mathrm{i} \lambda
\varphi_{p}\right).
$$

Let us consider massive Dirac particles with arbitrary
polarization vector $s_p$. After evaluations we obtain, that the
decomposition coefficients for a massive fermion with momentum
$p$, an arbitrary polarization vector $s_p$ and mass $m_p $ can
be written as scalar products of tetrad and physical vectors
\begin{eqnarray}
&& \mathcal{M}_{\rho,\;\lambda _p}\left(b_A,p,s_p\right)= \frac
1{\sqrt{\left(\tilde{b}_{-1} \cdot \xi _1^p \right)}}\left.\Bigg[
\delta _{\lambda,\; -\rho}  \left\{ \delta_{A,-1}
\left(\tilde{b}_{-1} \cdot \xi
_1^p\right)+\delta_{A,1}\left(\tilde{n}_{-\lambda _p} \cdot \xi
_1^p\right) \right\} \right.+
\nonumber\\
&&\left. + \delta _{\lambda,\; \rho
}\left\{\delta_{A,1}Y_{-1,\;-\lambda_{p} }^{\xi _{-1}^{p},\;\xi
_{1}^{p}}+\frac{\delta_{A,-1}}{m_p} X_{-1,-\lambda_{p} }^{\xi
_{-1}^{p},\;\xi _{1}^{p}} \right\} \right]\; , \label{anpic9}
\end{eqnarray}
where the scalar functions $Y^{p,q}, X^{p,q}$ are determined by
Eqs.(\ref{anpic12new1})-(\ref{anpic12new2}).

For $BDKS$ polarization states with the polarization vector
(\ref{stp33aa}) the matrix element (\ref{anpic9}) has a compact
form
\begin{eqnarray}
&&\mathcal{M}_{\rho,\;\lambda}\left(b_A,\;p,\;s_{KS}\right)=
\delta_{\lambda,\;-\rho}\left[ \delta_{A,-1}\sqrt{\left(p \cdot
\tilde{b}_{-1}\right) }+ \delta_{A,1}\frac{\left(p \cdot
\tilde{n}_{-\lambda}\right)}{\sqrt{\left(p \cdot
\tilde{b}_{-1}\right)}}\right]+
\nonumber\\
&& +\delta_{\lambda,\rho}\;\delta_{A,1}\frac{m_p} {\sqrt{\left(p
\cdot \tilde{b}_{-1}\right)}}\;. \label{dc1}
\end{eqnarray}

The matrix element (\ref{anpic3}) with the antifermion can be
easily obtained with the help of Eq.(\ref{stp31}):
\begin{eqnarray}
&&\widetilde{\mathcal{M}}_{\rho,\;\lambda
_p}\left(b_{A}\;,\left[p\right]\right)= \bar{u}_{\rho}
\left(b_{A}\right)\upsilon_{\lambda_{p}} \left(p,\;s_p\right)=
\nonumber\\
&& =\rho \lambda_{p}\;\mathcal{M}_{\rho,\;-\lambda
_p}\left(b_{A}\;,\left[p\right]\right)\;. \label{anpic3af}
\end{eqnarray}

\subsection{Recursion relation}

With the help of completeness relation (\ref{pic10}) the
amplitude (\ref{anpic1}) with $Q=Q_2 Q_1$ is expressed as
combinations of the lower-order matrix element
\begin{eqnarray}
&& \mathcal{M}_{\lambda _p,\;\lambda _k}\left(
\left[p\right],\left[k\right];Q_2 Q_1\right)=
\nonumber\\
&& =\sum_{\sigma ,A =-1}^1 \mathcal{M}_{\lambda
_p,\;\sigma}\left(\left[p\right],b_{A};Q_2\right)
\mathcal{M}_{-\sigma,\; \lambda _k }
\left(b_{-A},\left[k\right];Q_1\right)\;. \label{anpic2}
\end{eqnarray}
This insertion allows us to ``cut'' fermion chain into pieces of
fermion chains with basis spinors $u_{\lambda}\left(b_{A}\right)$.
Hence our formalism enables to calculate the blocks of the
Feynman diagrams  and then to use them in the calculation as
scalar functions. All possible Feynman amplitudes can be built up
from a set of ``building'' blocks.

Let us consider  the matrix element (\ref{anpic1}) with an
operator
\begin{equation}\label{anpic13}
Z^{\left(n\right)}=Q_{n} Q_{n-1}\cdots Q_1 Q_0
\end{equation}
with $Q_{0}=I$. In the Eq.(\ref{anpic13}) all operators $Q_j$
have an identical mathematical expressions. Using
Eq.(\ref{anpic2}) we find that
\begin{eqnarray}
&& \mathcal{M}_{\lambda _p,\;\lambda _k}\left(
\left[p\right],\left[k\right];Z^{\left(n\right)}\right)\equiv
\mathcal{M}_{\lambda _p,\;\lambda _k}^{\left(n\right)}\left(
\left[p\right],\left[k\right]\right)=
\nonumber\\
&& =\sum_{\sigma ,A =-1}^1 \mathcal{M}_{\lambda
_p,\;\sigma}\left(\left[p\right],b_{A}\right)
\mathcal{M}_{-\sigma,\; \lambda _k }^{\left(n\right)}
\left(b_{-A},\left[k\right]\right)\;, \label{anpic14}
\end{eqnarray}
where matrix element $\mathcal{M}_{-\sigma, \;\lambda _k
}^{\left(n\right)} \left(b_{-A},\left[k\right]\right)$ can be
calculated with the help of recursion relation
\begin{eqnarray}
&& \mathcal{M}_{-\sigma,\; \lambda _k }^{\left(n\right)}
\left(b_{-A},\left[k\right]\right)=
\nonumber\\
&&=\sum_{\rho ,C =-1}^1 \Gamma^{A,\;C}_{\sigma,\;\rho}\left[Q_n
\right] \mathcal{M}_{-\rho,\; \lambda _k }^{\left(n-1\right)}
\left(b_{-C},\left[k\right]\right)\;. \label{anpic15a}
\end{eqnarray}

Once scalar functions $\mathcal{M}_{\rho,\; \lambda _k }
\left(b_{A},\left[k\right]\right)$ and
$\Gamma^{A,C}_{\sigma,\rho}\left[Q_j \right]$ are known  (see
Eqs.(\ref{anpic7}), (\ref{anpic9})-(\ref{dc1}) and
Eqs.(\ref{pic16})-(\ref{pic16a})), it is possible to evaluate the
higher order of $\mathcal{M}_{-\sigma, \lambda _k
}^{\left(n\right)} \left(b_{-A},\left[k\right]\right)$ with the
help of recursion relation (\ref{anpic15a}).

\section {Examples}

Consider the ``toy'' example
\begin{eqnarray}
&&\mathcal{M}_{\lambda _{p},\;\lambda _{k}}^{\left( n\right)
}\left(
p,\;s_{p},\;k,\;s_{k}\right)=  \nonumber   \\
&&=\bar{u}_{\lambda _{p}}\left(p,s_{p}\right) \slash{q}_{n}
\slash{q}_{n-1}\ldots \slash{q}_{1}u_{\lambda _{k}}\left(
k,s_{k}\right),\label{anpic25}
\end{eqnarray}
where $q_j$ are some arbitrary four-vectors.

Therefore, we have that
\begin{equation}
Z^{\left( n\right) }= \slash{q}_{n} \slash{q}_{n-1}\ldots
\slash{q}_{1}\;. \label{anpic26}
\end{equation}
Using the Eqs.(\ref{pic16})-(\ref{pic16a}), (\ref{anpic15a}) we
find that
\begin{eqnarray}
&&\mathcal{M}_{-\rho,\; \lambda _k }^{\left(j\right)}
\left(b_{-C},\left[k\right]\right) =\left(q_{j}\cdot
\tilde{b}_{-C}\right) \mathcal{M}_{\rho,\; \lambda _k
}^{\left(j-1\right)} \left(b_{C},\left[k\right]\right)-
\nonumber\\
&& -C\;\left(q_{j}\cdot \tilde{n}_{C \rho
}\right)\mathcal{M}_{\rho,\; \lambda _k }^{\left(j-1\right)}
\left(b_{-C},\left[k\right]\right)\; \label{anpic28}
\end{eqnarray}
and
\begin{eqnarray}
&&\mathcal{M}_{-\rho,\; \lambda _k }^{\left(j\right)}
\left(b_{-C},\left[k\right]\right)
=\mathcal{B}_{C_{j-k}^{\prime},\rho_{j-k}^{\prime}}\left[q_{j},\ldots
q_{j-k}\right] \mathcal{M}_{-\rho_{j-k}^{\prime},\; \lambda _k
}^{\left(j-k\right)}
\left(b_{-C_{j-k}^{\prime}},\left[k\right]\right)+
\nonumber\\
&& +C_{j-k}^{\prime}\mathcal{N}_{-C_{j-k}^{\prime},\;
\rho_{j-k}^{\prime}}\left[q_{j},\ldots
q_{j-k}\right]\mathcal{M}_{-\rho_{j-k}^{\prime},\; \lambda _k
}^{\left(j-k\right)}
\left(b_{C_{j-k}^{\prime}},\left[k\right]\right)\;,
\label{anpic28a}
\end{eqnarray}
where $C_{j-k}^{\prime}=\left(-1\right)^{j-k} C,
\rho_{j-k}^{\prime}=\left(-1\right)^{j-k}\rho, k<j$.

With the help of Eq.(\ref{anpic14}) we can obtain the recursion
formulas for calculation matrix element (\ref{anpic25}):
\begin{eqnarray}
&& \mathcal{M}_{\lambda _{p},\;\lambda _{k}}^{\left(j\right)
}\left(\left[p\right],\left[k\right]\right)=
\nonumber\\
&& =\sum_{\rho ,C =-1}^1 \mathcal{M}_{\lambda
_p,\;\rho}\left(\left[p\right],b_{C}\right) \left[\left(q_{j}\cdot
\tilde{b}_{-C}\right) \mathcal{M}_{\rho, \;\lambda _k
}^{\left(j-1\right)} \left(b_{C},\left[k\right]\right)\right.-
\nonumber\\
&& \left.-C\;\left(q_{j}\cdot \tilde{n}_{C \rho
}\right)\mathcal{M}_{\rho,\; \lambda _k }^{\left(j-1\right)}
\left(b_{-C},\left[k\right]\right) \right]\;, \label{anpic27}
\end{eqnarray}
and
\begin{eqnarray}
&& \mathcal{M}_{\lambda _{p},\;\lambda _{k}}^{\left(j\right)
}\left(\left[p\right],\left[k\right]\right)=
\nonumber\\
&& =\sum_{\rho ,C =-1}^1 \mathcal{M}_{\lambda
_p,\;\rho}\left(\left[p\right],\;b_{C}\right)\left[\right.
\mathcal{B}_{C_{j-k}^{\prime},\rho_{j-k}^{\prime}}\left[q_{j},\ldots
q_{j-k}\right] \mathcal{M}_{-\rho_{j-k}^{\prime},\; \lambda _k
}^{\left(j-k\right)}
\left(b_{-C_{j-k}^{\prime}},\left[k\right]\right)+
\nonumber\\
&& +C_{j-k}^{\prime}\mathcal{N}_{-C_{j-k}^{\prime},\;
\rho_{j-k}^{\prime}}\left[q_{j},\ldots
q_{j-k}\right]\mathcal{M}_{-\rho_{j-k}^{\prime},\; \lambda _k
}^{\left(j-k\right)}
\left(b_{C_{j-k}^{\prime}},\left[k\right]\right)\left.\right]\;.
\label{anpic27a}
\end{eqnarray}
We have obtained  that the matrix element (\ref{anpic25})  can be
represented as a combination of the scalar functions
$\mathcal{B},\mathcal{N}$, decomposition coefficients
$\mathcal{M}_{\lambda _p,\rho}\left(\left[p\right],b_{C}\right)$
and lower order matrix elements.

\subsection{Photon emission}
Let us consider the matrix element where photon with momentum $k$
and helicity $\sigma=\pm 1$ emission occurs from the incoming
electron
\begin{eqnarray}
&& \mathcal{M}_{\lambda _2,\;\lambda _1} \left(
\left[p_2\right],\left[p_1\right]; Q\;Z_{k} \right) =
\bar{u}_{\lambda _{2}}\left( p_2,\;s_{p_2}\right) Q\;Z_{k}
\;u_{\lambda _1}\left(p_1,\;s_{p_1}\right) ~~\mbox{with}
\label{fot1}\\
&&Z_k=e\;
\frac{\left(\slash{p}_{1}-\slash{k}+m\right)\slash{\varepsilon
}_{\sigma}\left(k\right)}{\left(p_{1}-k\right)^{2}-m^{2}}
\;.\label{fot2}
\end{eqnarray}
Using the algebra of $\gamma$-matrix and Dirac equation the
operator $Z_k$ can be rewritten as
\begin{equation}\label{fot3}
Z_k=-e\;\left\{\frac{\left(p_1\;\varepsilon_{\sigma}\left(k\right)\right)}{\left(p_1\;k
\right)}-\frac{\slash{k}\slash{\varepsilon
}_{\sigma}\left(k\right)}{2\left(p_1\;k \right)}\right\}\;.
\end{equation}
Now we get
\begin{eqnarray}
&& \mathcal{M}_{\lambda _2,\;\lambda _1} \left(
\left[p_2\right],\left[p_1\right]; Q\;Z_{k} \right)
=-e\;\frac{\left(p_1\;\varepsilon_{\sigma}\left(k\right)\right)}{\left(p_1\;k
\right)}\mathcal{M}_{\lambda _2,\;\lambda _1} \left(
\left[p_2\right],\left[p_1\right]; Q \right)+
\nonumber\\
&&+\frac{e}{2\left(p_1\;k \right)}\mathcal{M}_{\lambda
_2,\;\lambda _1}\left( \left[p_2\right],\left[p_1\right]; Q
\;\slash{k}\slash{\varepsilon }_{\sigma}\left(k\right)\right)\;.
\label{fot4}
\end{eqnarray}
The recursion technique (see Eq.(\ref{anpic2})) imply
\begin{eqnarray}
&& \mathcal{M}_{\lambda _2,\;\lambda _1}\left(
\left[p_2\right],\left[p_1\right]; Q \;\slash{k}\slash{\varepsilon
}_{\sigma}\left(k\right)\right)=
\nonumber\\
&&=\sum_{\sigma ,C =-1}^1 \mathcal{M}_{\lambda
_2,\;\sigma}\left(\left[p_2\right],b_{C};Q\right)
\mathcal{M}_{-\sigma, \lambda _1 } \left(b_{-C},\left[p_1\right];
\slash{k}\slash{\varepsilon }_{\sigma}\left(k\right)\right)=
\nonumber\\
&&=\sum_{\sigma ,C =-1}^1 \mathcal{M}_{\lambda
_2,\;\sigma}\left(\left[p_2\right],b_{C};Q\right)
\Gamma_{\sigma,-\lambda _1}^{C,-1}
\left[\slash{k},\slash{\varepsilon}_{\sigma}\left(k\right),T_{\lambda_1}\left(p_1,
s_{p_1}\right)\right]\;,\label{fot5}
\end{eqnarray}
where
\begin{equation}\label{toper1}
T_{\lambda}\left(p, s_{p}\right) =\frac{ \slash{\xi}^{p}_{1}+
\slash{\xi}^{p}_{-1}\slash{\xi}^{p}_{1}/ m_p}{\sqrt{\left(
\tilde{b}_{-1}\cdot \xi^{p}_{1}\right)}}\;
\end{equation}
for fermion with arbitrary polarization vector (see
Eq.(\ref{anpic8})),
\begin{equation}\label{toper2}
T_{\lambda}\left(p, s_{p}\right) =\frac{
\slash{p}+m_p}{\sqrt{\left(p \cdot \tilde{b}_{-1}\right)}}\;
\end{equation}
for fermion with $BDKS$ polarization vector (\ref{stp34aa}),
\begin{equation}\label{toper3}
T_{\lambda}\left(p, s_{p}\right)=\frac{ \slash{p}}{\sqrt{\left(p
\cdot \tilde{b}_{-1}\right)}} \;
\end{equation}
for massless fermion (see Eq.(\ref{anpic6})).

Let us consider initial massive fermion with $BDKS$ polarization
vector in expression (\ref{fot5}). After calculations, for the
matrix element  (\ref{fot1}) with the $BDKS$ polarization state
of initial fermion  and  helicity $\sigma$ of photon we have the
exact formula in terms of lower-order matrix elements with
operator $Q$ and scalar functions $\mathcal{B},\mathcal{N}$:
\begin{eqnarray}
&& \mathcal{M}_{\lambda _2,\;\lambda _1}\left(
\left[p_2\right],\left[p_1\right]; Q \;Z_k\right)=
-e\;\frac{\left(p_1\;\varepsilon_{\sigma}\left(k\right)\right)}{\left(p_1\;k
\right)}\mathcal{M}_{\lambda _2,\;\lambda _1} \left(
\left[p_2\right],\left[p_1\right]; Q \right)+
\nonumber\\
&&+\frac{e}{2\left(p_1\;k
\right)\sqrt{\left(p_{1}\;\tilde{b}_{-1}\right)}} \bigg[
\mathcal{M}_{\lambda_2,\;\lambda_1}
\left(\left[p_2\right],b_{-1};Q\right)
\mathcal{N}^{~k,~\varepsilon_{\sigma}\left(k\right),\;p_{1}}_{-1,-\lambda_{1}}+
\nonumber\\
&&+ \mathcal{M}_{\lambda
_2,\;\lambda_1}\left(\left[p_2\right],b_{1};Q\right)
\mathcal{B}^{~k,~\varepsilon_{\sigma}\left(k\right),\;p_{1}}_{-1,-\lambda_{1}}
 + m_{p_1}\left\{ \right. \mathcal{M}_{\lambda_2,\;-\lambda_1}
\left(\left[p_2\right],b_{-1};Q\right)
Y^{~k,~\varepsilon_{\sigma}\left(k\right)}_{-1,-\lambda_{1}}
\nonumber\\
&& +\mathcal{M}_{\lambda
_2,\;-\lambda_1}\left(\left[p_2\right],b_{1};Q\right)
X^{~k,~\varepsilon_{\sigma}\left(k\right)}_{-1,-\lambda_{1}}
\left.\right\}\bigg] \;. \label{fot5a}
\end{eqnarray}
Scalar functions
$\mathcal{B}^{~k,~\varepsilon_{\sigma}\left(k\right),p_{1}}_{A,\;\lambda}$,
$\mathcal{N}^{~k,~\varepsilon_{\sigma}\left(k\right),p_{1}}_{A,\;\lambda}$
are determined in terms of scalar product with the help of  the
Eqs.(\ref{bn3})-(\ref{bn4}) and scalar functions
$X^{~k,~\varepsilon_{\sigma}\left(k\right)}_{-1,-\lambda_{1}}$,
$Y^{~k,~\varepsilon_{\sigma}\left(k\right)}_{-1,-\lambda_{1}}$
are determined by Eqs.(\ref{anpic12new1})-(\ref{anpic12new2}).

Scalar functions can be easily calculated in terms of physical
vector components. Using a vector of polarization of a photon as
Eq.(\ref{picc1b}) we obtain simple result:
\begin{eqnarray}
&&
\mathcal{B}^{~k,~\varepsilon_{\sigma}\left(k\right),\;p}_{-1,\;-\lambda_1}=
\sqrt{2}\;\sigma\left(p^{-}k_{T,\;\lambda_{1}}
\left[\delta_{\lambda_{1},\;\sigma}-1\right]+k^{-}p_{T,\;-\sigma}\right)\;,
\nonumber\\
&&
\mathcal{N}^{~k,~\varepsilon_{\sigma}\left(k\right),\;p}_{-1,\;-\lambda_1}=
\frac{\sqrt{2}}{k^{-}}\left[p^{-} k_{T,-\lambda_{1}}k_{T,-\sigma}
+k^{-}\left(\delta_{\lambda_{1},\;-\sigma}k^{+}p^{-}-
\delta_{\lambda_{1},\;\sigma} p_{T,\;-\sigma}
k_{T,\;-\sigma}\right) \right]\;,
\nonumber\\
&&Y^{~k,~\varepsilon_{\sigma}\left(k\right)}_{-1,\;-\lambda_{1}}=
-\sqrt{2}\;\lambda_{1}\;\delta_{\lambda_{1},\;-\sigma}
k_{T,\;-\sigma}\;,\;~~X^{~k,~\varepsilon_{\sigma}\left(k\right)}_{-1,\;-\lambda_{1}}=
-\sqrt{2}\;\delta_{\lambda_{1},\;-\sigma}\;k^{-}\;,
\label{scalarfunct}
\end{eqnarray}
where
\begin{equation}
p^{\pm }=p^0 \pm p^z\;,\;~ p_{T,\;\lambda}=p^x+\mathrm{i}\;\lambda
p^y\;. \label{index1}
\end{equation}

\subsection{The process $e^{+} e^{-} \to n \gamma $}

Consider the process
\begin{equation}\label{anpic29}
e^{+}\left(p_{2}, \sigma_{2}\right)+ e^{-}\left(p_{1},
\sigma_{1}\right) \to \gamma \left(k_1,\lambda_{1}\right)+\gamma
\left(k_2,\lambda_{2}\right)+ \cdots +\gamma
\left(k_n,\lambda_{n}\right),
\end{equation}
where the momenta of the particles and spin numbers are given
between parentheses.

The Feynman diagrams of the processes (\ref{anpic29}) contain the
matrix element
\begin{eqnarray}
&&M_{\sigma _{2},\;\sigma _{1}}^{\left( \lambda _{1},\;\lambda
_{2}, \ldots \lambda _{n}\right) }\left(
p_{2},s_{p_{2}},p_{1},s_{p_{1}};k_{1},k_{2},\ldots k_{n}\right)=
\nonumber \\
&&={M}_{\sigma _{2},\;\sigma _{1}}^{\left(n\right) }\left(
\left[p_2\right],\left[p_1\right]\right)= \bar{\upsilon }_{\sigma
_{2}}\left( p_{2},s_{p_{2}}\right)
\slash{ \varepsilon }_{\lambda _{n}}\left( k_{n}\right) \cdots \nonumber \\
&& \cdots \slash{\varepsilon } _{\lambda _{3}}\left( k_{3}\right)
\frac{\slash{Q}_{2}+m}{Q_{2}^{2}-m^{2}} \slash{\varepsilon
}_{\lambda _{2}}\left( k_{2}\right) \frac{\slash{Q}
_{1}+m}{Q_{1}^{2}-m^{2}}\slash{\varepsilon }_{\lambda _{1}}\left(
k_{1}\right) u_{\sigma _{1}}\left( p_{1},s_{p_{1}}\right)+\nonumber \\
&&+\left(n!-1 \right)\mbox{other permutations of}
\left(1,2,\ldots, n\right)\;, \label{anpic20}
\end{eqnarray}
where
\begin{equation}
Q_{j}=p_{1}-\sum_{i=1}^{j}k_{i}\; . \label{anpic21}
\end{equation}
Hence, we have that
\begin{equation}
Z^{\left( n\right)}=\frac{\slash{Q}_{n}+m}{Q_{n}^{2}-m^{2}}\slash{
\varepsilon }_{\lambda _{n}}\left( k_{n}\right) \cdots
\frac{\slash{Q} _{1}+m}{Q_{1}^{2}-m^{2}}\slash{\varepsilon
}_{\lambda _{1}}\left( k_{1}\right) \label{anpic22}
\end{equation}
and
\begin{eqnarray}
&& {M}_{\sigma _{2},\sigma _{1}}^{\left(n\right)
}\left(\left[p_2\right],\left[p_1\right] \right)
\nonumber\\
&& =\sum_{\rho ,C =-1}^1 \widetilde{\mathcal{M}}_{\sigma
_{2},\rho}\left(\left[p_{2} \right],b_C;
\slash{\varepsilon}_{\lambda _{n}}\left(k_{n}\right)\right)
\mathcal{M}_{-\rho,\sigma _{1}}\left(b_{-C},\left[p_{1}\right];
Z^{\left(n-1\right)} \right) \label{anpic23}\;.
\end{eqnarray}
Here
\begin{equation}
\widetilde{\mathcal{M}}_{\sigma _{2},\rho}\left(\left[p_{2}
\right],b_C; \slash{\varepsilon}_{\lambda
_{n}}\left(k_{n}\right)\right) =\bar{\upsilon }_{\sigma
_{2}}\left( p_{2},s_{p_{2}}\right)\slash{\varepsilon}_{\lambda
_{n}}\left(k_{n}\right)  u_{\rho}\left(b_{C}\right)
\label{anpic23a}
\end{equation}
and
\begin{equation}
\mathcal{M}^{\left(n-1\right)}_{-\rho,\sigma
_{1}}\left(b_{-C},\left[p_{1}\right]\right)=\mathcal{M}_{-\rho,\sigma
_{1}}\left(b_{-C},\left[p_{1}\right]; Z^{\left(n-1\right)}
\right) \;. \label{anpic23abc}
\end{equation}

Using the expressions (\ref{pic16a}) and (\ref{picc1b}) we
obtain, that (\ref{anpic23a}) for $BDKS$ massive Dirac spinor
(\ref{stp34aa}) is determined by
\begin{eqnarray}
&&\widetilde{\mathcal{M}}_{\sigma _{2},\rho}\left(\left[p_{2}
\right],b_C; \slash{\varepsilon}_{\lambda
_{n}}\left(k_{n}\right)\right)=
\nonumber\\
&& =1/\sqrt{\left(p_2 \cdot \tilde{b}_{-1} \right)}
\left.\bigg[\delta_{\sigma_{2},\;\rho}
\left(\delta_{C,-1}X_{-1,\;\sigma_{2}}^{p_2,\;\varepsilon_{n}}+
\delta_{C,1}Y_{1,\;\sigma_{2}}^{p_2,\;\varepsilon_{n}}\right)\right.
\nonumber\\
&&\left. -m\;
\delta_{\sigma_{2},\;-\rho}\left(\delta_{C,-1}\left(\tilde{b}_{-1}
\cdot
\varepsilon_{n}\right)-\delta_{C,1}\left(\tilde{n}_{\sigma_{2}}
\cdot \varepsilon_{n}\right)\right)\right] \label{lastmn}
\end{eqnarray}
with $\varepsilon_{n}=\varepsilon_{\lambda _{n}}\left(
k_{n}\right)$.

The final recursion relation of the process (\ref{anpic29}) with
arbitrary helicities of the photons and $BDKS$ polarization
states of positron is written as
\begin{eqnarray}
&& {M}_{\sigma _{2},\sigma _{1}}^{\left(n\right)
}\left(\left[p_2\right],\left[p_1\right] \right)
=1/\sqrt{\left(p_2 \cdot \tilde{b}_{-1} \right)}
\nonumber\\
&& \left[X_{-1,\;\sigma_{2}}^{p_2,\;\varepsilon_{n}}
\mathcal{M}_{-\sigma_{2},\sigma
_{1}}^{\left(n-1\right)}\left(b_{1}, \left[p_{1}\right]\right)+
Y_{1,\;\sigma_{2}}^{p_2,\;\varepsilon_{n}}\mathcal{M}_{-\sigma_{2},\sigma
_{1}}^{\left(n-1\right)}\left(b_{-1}, \left[p_{1}\right]\right)
\right.\nonumber\\
&& \left.+m\;\left(\left(\tilde{n}_{\sigma_{2}} \cdot
\varepsilon_{n}\right)\mathcal{M}_{\sigma_{2},\sigma
_{1}}^{\left(n-1\right)}\left(b_{-1}, \left[p_{1}\right]\right)+
\left(\tilde{b}_{-1} \cdot
\varepsilon_{n}\right)\mathcal{M}_{\sigma_{2},\sigma
_{1}}^{\left(n-1\right)}\left(b_{1},
\left[p_{1}\right]\right)\right)\right] \;, \label{finalres}
\end{eqnarray}
where (see Eq.(\ref{anpic27}) and Eq.(\ref{anpic27a})) the matrix
element $\mathcal{M}_{-\rho ,\sigma _{1}}^{\left( j\right) }\left(
b_{-C},\left[ p_{1}\right] \right)$ with arbitrary polarization
vector of electron is calculated by means of the recursion formula
\begin{eqnarray}
&&\mathcal{M}_{-\rho ,\sigma _{1}}^{\left( j\right) }\left(
b_{-C},\left[
p_{1}\right] \right) =\frac{1}{Q_{j}^{2}-m^{2}}\left\{ {}\right.   \nonumber \\
&&m\left[ {}\right. \left( \varepsilon _{j}\cdot b_{-C}\right)
\mathcal{M} _{-\rho ,\sigma _{1}}^{\left( j-1\right) }\left(
b_{-C},\left[ p_{1}\right] \right) -C\;\left( \varepsilon
_{j}\cdot n_{C\rho }\right) \mathcal{M} _{-\rho ,\sigma
_{1}}^{\left( j-1\right) }\left( b_{C},\left[ k\right]
\right) \left. {}\right] +  \nonumber \\
&&+Y_{C,\rho }^{Q_{j},\varepsilon _{j}}\mathcal{M}_{\rho ,\sigma
_{1}}^{\left( j-1\right) }\left( b_{C},\left[ p_{1}\right] \right)
+C\;X_{-C,\rho }^{Q_{j},\;\varepsilon _{j}}\mathcal{M}_{\rho
,\sigma _{1}}^{\left( j-1\right) }\left( b_{-C},\left[
p_{1}\right] \right) \left. {}\right\} \;. \label{anpic24x}
\end{eqnarray}

\section{Summary and Acknowledgements}
In present paper we have formulated a new effective method to
calculate the Feynman amplitudes for various processes with
fermions of arbitrary polarizations. In our method it is much
easier to keep track of partial results and to set up recursive
schemes of evaluation which compute and store for later use
subdiagrams of increasing size and complexity.

In our approach of the matrix element calculation:
\begin{description}
\item[1] we don't use  an explicit form of Dirac spinors and
$\gamma$--matrices
\item[2] we don't use the calculation of traces
\item[3] as well as in the trace methods the matrix element
of Feynman amplitudes is reduced to the combination of scalar
products of momenta and polarization vectors.
\item[4] Unlike spinor technique in different variants
\cite{Berendz}-\cite{Zhan} in this method we don't use either
Chisholm identities, or the presentation of the contraction
$\slash{p}$ with four vector $p$ and of the polarization vector
of bosons through the Dirac spinors.
\item[5] Unlike \texttt{WvD} technique
\cite{Giele},\cite{Dittmaier} in this method we don't use special
Feynman rules for calculating of the matrix elements.
\item[6] Expression for matrix element  calculates $\mathcal{M}_{\lambda _p,\lambda _k} \left(p,s_p,\; k,s_k\;
;Q\right)$ (\ref{anpic1}) for all values $\lambda _p, \lambda _k$
simultaneously.
\end{description}
The  recursive algorithms can be easily realized in the various
systems of symbolic calculation (Mathematica, Maple, Reduce,
Form) and in such packages as \texttt{FeynArts} \cite{FeynArts},
\texttt{FeynCalc} \cite{FeynCalc}, \texttt{HIP} \cite{hip} and so
on.

I would like to thank  organizers for warm an kind hospitality
throughout the Conference. Also I want to thank A.L. Bondarev for
his useful remarks and discussion.

\end{document}